\documentclass{article}
\usepackage{spconf,amsmath,graphicx}
\usepackage{algorithm}
\usepackage{algorithmic}
\usepackage{cite}
\usepackage{amsfonts}
\usepackage{fancyhdr}
\usepackage{amsmath,amssymb,amsthm}%
\usepackage{graphicx}%
\usepackage{listings}%
\usepackage{arydshln}
\usepackage{subfig}%
\usepackage{algorithmic}%
\newtheorem{mylem}{Lemma}
\newtheorem{myrem}{Remark}%

\title{Position and Orientation Estimation of a Rigid Body: \\Rigid Body Localization}
\name{Sundeep Prabhakar Chepuri, Geert Leus, and Alle-Jan van der Veen\thanks{This work was supported in part by STW under the FASTCOM project (10551) and in part by NWO-STW under the VICI program (10382). }}
\address{Faculty of Electrical Engineering, Mathematics, and Computer Science (EEMCS)\\
Delft University of Technology (TU Delft), The Netherlands\\
Email:~\{s.p.chepuri; g.j.t.leus; a.j.vanderveen\}@tudelft.nl.}
\begin{document}
\ninept
\maketitle
\begin{abstract}
Rigid body localization refers to a problem of estimating the position of a rigid body along with its orientation using anchors. We consider a setup in which a few sensors are mounted on a rigid body. The absolute position of the rigid body is not known, but, the relative position of the sensors or the topology of the sensors on the rigid body is known. We express the absolute position of the sensors as an affine function of the Stiefel manifold and propose a simple least-squares (LS) estimator as well as a constrained total least-squares (CTLS) estimator to jointly estimate the orientation and the position of the rigid body. To account for the perturbations of the sensors, we also propose a constrained total least-squares (CTLS) estimator. Analytical closed-form solutions for the proposed estimators are provided. Simulations are used to corroborate and analyze the performance of the proposed estimators.\end{abstract}
\begin{keywords}
Rigid body localization, Stiefel manifold, attitude estimation, tilt estimation, sensor networks.
\end{keywords}
\section{Introduction}
Advances in wireless sensor technology and their usage in networks have given birth to a variety of sensing, monitoring, and control applications. The majority of  applications with a wireless sensor network (WSN) rely on two fundamental aspects: distributed data sampling and information fusion. For the data to be meaningful it is important to know not only the time instance (temporal information) at which the data is acquired, but also the location (spatial information) where the data is acquired. Identifying the sensor's location  is a well-studied topic~\cite{Gusta05SPM}, and is commonly referred to as \textit{localization}. 

Localization can be either absolute or relative. In absolute localization, the nodes are usually localized using a few reference nodes whose positions are known. Absolute localization problems are typically solved using range-square methods from measurements based on certain physical phenomena, e.g., time-of-arrival (TOA)~\cite{Gusta05SPM,localizationSPM}.  Localization can also be relative. In relative localization, the aim is to identify the topology of the network, and determining the location of the nodes relative to other nodes is sufficient. Classical solutions to relative localization are based on multi-dimensional scaling (MDS)~\cite{MDS,MDShadi}.

In this paper, we provide a new and different flavor of localization, called \textit{rigid body localization}. The problem in rigid body localization is to identify the location of the body in a three-dimensional space and also the orientation of the body along these three-dimensions. Rigid body localization has a huge potential in a variety of different fields. To list a few, it is useful in the areas of underwater (or in-liquid) systems, orbiting satellites, mechatronic systems, unmanned aircrafts, gaming consoles, or automobiles.  In such applications, traditional localization of the node(s) is not sufficient. For example, in an autonomous underwater vehicle (AUV)~\cite{underwaterAV}, or an orbiting satellite~\cite{olfar}, the sensing platform is not only subject to motion but also to rotation. In such cases, together with positioning, determining the orientation of the body also forms a key component, and is essential for both controlling (maneuvering) and monitoring purposes.  

Commonly the term \textit{attitude estimation} (for flights and spacecrafts) or \textit{tilt sensing} (for industrial equipments and consumer devices) is used for determining the orientation of the object in a three-dimensional space which typically uses inertial sensors~\cite{GPSattitude}, or accelerometers~\cite{accelerometer}. However, inertial sensors and accelerometers generally suffer from drift errors. On the other hand, in rigid body localization we propose to exploit the communication packets containing the ranging information, just as in traditional localization schemes~\cite{Gusta05SPM}, to estimate both the rotations and the translations. In short, we present rigid body localization as an estimation problem from a signal processing perspective.

More specifically, we consider a rigid body on which a few sensors are mounted. The absolute location of the rigid body itself is unknown, but, the relative position of the sensors or the sensor topology on the rigid body is known. A novel problem to jointly position the rigid body and estimate its orientation using a few nodes with known absolute locations (anchors) is considered. For this purpose, we parameterize the Stiefel manifold~\cite{Stiefel1999} with a known sensor topology and propose a new \textit{least-squares} (LS) estimator and also a \textit{constrained least-squares} (CLS) estimator. The sensor positions are usually perturbed during fabrication of the body or if the body is not entirely rigid. To take these perturbations into account, we also propose a \textit{constrained total least-squares} (CTLS) estimator.  Analytical closed-form solutions for the proposed estimators are provided. Simulations are provided to validate and analyze the performance of the proposed estimators.

{{\textsl{Notation}: Upper (lower) bold face letters are used for matrices (column vectors); 
$(\cdot)^T$ denotes transposition; $\mathrm{diag}(.)$ refers to a block diagonal matrix with the matrices in its argument on the main diagonal; $\mathbf{1}_N$ ($\mathbf{0}_N$) denotes the $N \times 1$ vector of ones (zeros); $\small \mathbf{I}_N$ is an identity matrix of size $N$; $\mathbb{E}(.)$ denotes the expectation operation; $\otimes$ is the Kronecker product;  $(.)^\dag$ denotes the pseudo inverse, i.e., for a full column-rank matrix ${\bf A}$ the pseudo inverse is given by ${\bf A}^\dag = ({\bf A}^T{\bf A}^{-1}){\bf A}^T$;
$\mathrm{vec}(.)$ is a vector formed by stacking the columns of its matrix argument; $\mathrm{vec}^{-1}(.)$ is a matrix formed by the inverse $\mathrm{vec}(.)$ operation.

\section{Problem formulation}
\subsection{The model and preliminaries} \label{sec:model}
\begin{figure} [!t]
\centering
\includegraphics[width=\columnwidth]{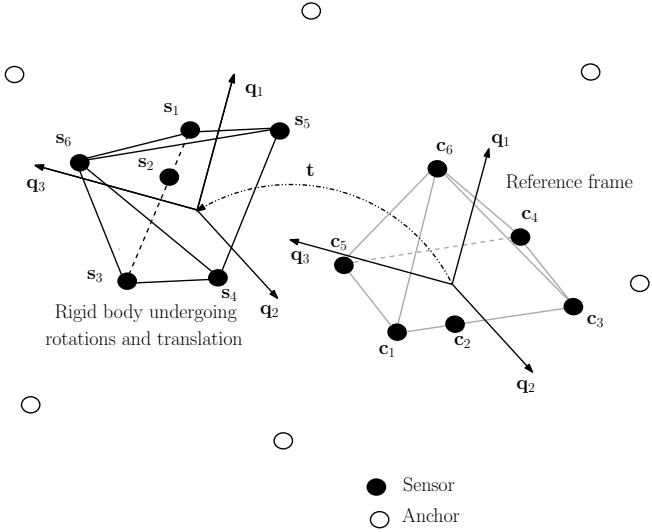}
\caption{An illustration of the sensors on a rigid body undergoing a rotation and a translation.}
\label{fig:model}
\vspace*{-5mm}
\end{figure}
Consider a network with $M$ anchors (nodes with known absolute locations) and $N$ sensors in a $3$-dimensional space. The sensors are mounted on a rigid body (e.g., a plane or a pyramid) as illustrated in Fig.~\ref{fig:model}. The relative position of these sensors or its topology on the rigid body is known up to a certain accuracy. However, the absolute position of the sensors or the rigid body itself in the $3$-dimensional space is not known. The rigid body experiences rotations and translations in each dimension.  

The sensors are mounted on the rigid body (e.g., in the factory) and the topology of how these sensors are mounted is known. In other words, we can connect a so-called {\it reference frame} to the rigid body, as illustrated in Fig. 1, and in that reference frame, the coordinates of the $n$th sensor are given by the known $3 \times 1$ vector ${\bf c}_n= [ c_{n,1}, c_{n,2}, c_{n,3} ]^T$. So the sensor topology is basically determined by the matrix ${\bf C} = [ {\bf c}_1, {\bf c}_2, \dots, {\bf c}_N ] \in {\mathbb R}^{3 \times N}$. Let the absolute coordinates of the $m$th anchor and the $n$th sensor be denoted by a $3 \times 1$ vector ${\bf a}_m$ and ${\bf s}_n$, respectively. These absolute positions of the anchors and the sensors are collected in the matrices ${\bf A} = [{\bf a}_1,{\bf a}_2,\ldots,{\bf a}_M] \in \mathbb{R}^{3 \times M}$ and ${\bf S} = [{\bf s}_1,{\bf s}_2,\ldots,{\bf s}_N] \in \mathbb{R}^{3 \times N}$, respectively.

The pairwise distance between the $m$th anchor and the $n$th sensor is denoted by ${r}({\bf a}_m,{\bf s}_n)= {\|{\bf a}_m-{\bf s}_n\|}_2$ and is typically obtained from ranging~\cite{Gusta05SPM,localizationSPM,ChepuriSPL}. The range measurements can be expressed as
\begin{equation}
\begin{aligned}
\label{eq:range}
 \hat{r}({\bf a}_m,{\bf s}_n) &={r}({\bf a}_m,{\bf s}_n) + e_{mn} \\
 \end{aligned}
\end{equation} where $e_{m,n}$ is the additive stochastic noise resulting from the ranging process.  Assuming TOA-based ranging, we model $e_{mn}$ as an i.i.d. zero mean white random process with a variance $\sigma^2(e_{mn}) = \frac{3c^2}{N_s\kappa}\frac{{r}^2({\bf a}_m,{\bf s}_n)}{\mathrm{SNR}}$~\cite{yiyineurasip}. Here, $c$ is the speed of a wave in a medium, $N_s$ is the number of samples used in the ranging process, $\kappa$ is a constant, and ${\mathrm{SNR}}$ is the signal-to-noise ratio of the range measurements. The ${r}^2({\bf a}_m,{\bf s}_n)$ term in the variance is due to the path-loss model assumption and penalizes the range measurements based on distance. Since all the sensors are mounted on the rigid body, it is reasonable to assume that all sensors experience approximately the same path-loss, especially when the anchors are far away from the rigid body. Hence, we use a simplified noise model with variance 
\begin{equation}
\begin{aligned}
\label{eq:variance_approx}
\sigma^2(e_{mn}) \approx  \frac{3c^2}{N_s\kappa}\frac{{r}^2({\bf a}_m,{\bf s}_1)}{\mathrm{SNR}}. 
\end{aligned}
\end{equation} Here, we choose sensor ${\bf s}_1$ just for illustration purposes, and in principle, this can be any sensor.

We can now write the squared pairwise distance between the $m$th anchor and the $n$th sensor as
\begin{equation}
\begin{aligned}
\label{eq:rangesquare}
 d({\bf a}_m,{\bf s}_n) = {r}^2({\bf a}_m,{\bf s}_n) = {\|{\bf a}_m\|^2-2{\bf a}_m^T{\bf s}_n+\|{\bf s}_n\|^2}
\end{aligned}
\end{equation} and
\begin{equation}
\begin{aligned}
\label{eq:rangesquare}
 \hat{d}({\bf a}_m,{\bf s}_n) = \hat{r}^2({\bf a}_m,{\bf s}_n) &= {r}^2({\bf a}_m,{\bf s}_n) + 2{r}({\bf a}_m,{\bf s}_n)e_{m,n}+e_{m,n}^2\\
 &=  {d}({\bf a}_m,{\bf s}_n) +n_{m,n}
\end{aligned}
\end{equation}
where $n_{m,n} = 2{r}({\bf a}_m,{\bf s}_1)e_{m,n} + e_{m,n}^2$ is the new noise term obtained due to squaring. We can compute the mean $\mathbb{E}(n_{m,n}) \approx 0$ and the variance $\sigma^2_m  = \mathbb{E}(n_{m,n}^2) \approx 4\sigma^2(e_{m,n}){r}^2({\bf a}_m,{\bf s}_1)$, ignoring the higher-order terms under the condition of sufficiently small errors.

Defining the $M \times 1$ vectors ${\bf d}({\bf s}_n) = [\hat d({\bf a}_1,{\bf s}_n),\ldots,\hat d({\bf a}_M,{\bf s}_n)]^T$ and ${\bf u} = [\|{\bf a}_1\|^2,\|{\bf a}_2\|^2,\ldots,\|{\bf a}_M\|^2]^T$, we can now write the squared pairwise distances of the $n$th sensor to each anchor in a vector form as
\begin{equation}
\label{eq:sqpairwisedistance}
{\bf d}({\bf s}_n) = {\bf u} - 2{\bf A}^T{\bf s}_n + \|{\bf s}_n\|^2 {\bf 1}_M + {\bf n}({\bf s}_n)
\end{equation} where ${\bf n}({\bf s}_n) = [n_{1,n},n_{2,n},\ldots,n_{M,n}]^T \in \mathbb{R}^{M \times 1}$ is the error vector. The $M \times M$ covariance matrix of the error vector ${\bf n}({\bf s}_n)$ will be ${\bf \Sigma}_n = \mathrm{diag}(\sigma_1^2,\sigma_2^2,\ldots,\sigma_M^2)$. We whiten (\ref{eq:sqpairwisedistance}) to obtain an identity noise covariance matrix by multiplying both sides of (\ref{eq:sqpairwisedistance}) with ${\bf W} \in \mathbb{R}^{M \times M}$, which leads to
\begin{equation}
\label{eq:sqpairwisedistance1}
{\bf W} {\bf d}({\bf s}_n) = {\bf W} ({\bf u} - 2{\bf A}^T{\bf s}_n + \|{\bf s}_n\|^2 {\bf 1}_M + {\bf n}({\bf s}_n))
\end{equation} 
The optimal ${\bf W}$ is ${\bf W}^* = \boldsymbol{\Sigma}_n^{-1/2}$ but depends on the unknown $r({\bf a}_m, {\bf s}_1)$. Hence, we use ${\bf W} = \hat{\boldsymbol{\Sigma}}_n^{-1/2}$, where  $\hat{\boldsymbol{\Sigma}}_n^{-1/2}$ is computed using $\hat{r}({\bf a}_m,{\bf s}_1)$.

In order to eliminate $\|{\bf s}_n\|^2$ and thus the vector ${\bf W} {\bf 1}_M$, the conventional technique is to apply an orthogonal projection matrix ${\bf P}_M \triangleq {\bf I}_M - \frac{{\bf W}{\bf 1}_M{\bf 1}_M^T{\bf W}}{{\bf 1}_M^T{\bf W}{\bf W}{\bf 1}_M} \in \mathbb{R}^{M \times M}$, such that ${\bf P}_M{\bf W}{\bf 1}_M ={\bf 0}$. However, this would again color the noise. To avoid this, we propose to use a unitary decomposition of ${\bf P}_M$, i.e.,  ${\bf P}_M = {\bf U}_M{\bf U}_M^T$ where ${\bf U}_M$ is a $M \times (M-1)$ matrix obtained by collecting orthonormal basis vectors of the null-space of ${\bf W}{\bf 1}_M$ so that ${\bf U}_M^T{\bf W}{\bf 1}_M={\bf 0}$.

In order to eliminate the $\|{\bf s}_n\|^2{\bf W}{\bf 1}_M$ term in (\ref{eq:sqpairwisedistance1}) without coloring the noise, we left-multiply both sides of (\ref{eq:sqpairwisedistance}) with ${\bf U}_M^T$, which leads to 
 \begin{equation}
\begin{aligned}
{\bf U}_M^T{\bf W} ({\bf d}({\bf s}_n)-{\bf u}) =& -2{\bf U}_M^T{\bf W}{\bf A}^T{\bf s}_n 
+ {\bf U}_M^T{\bf W}{\bf n}({\bf s}_n),
\end{aligned}
\label{eq:linear}
\end{equation} 
We can now stack (\ref{eq:linear}) for all the $N$ sensors as
\begin{equation}
{\bf U}_M^T {\bf W} {\bf D}= -2{\bf U}_M^T{\bf W}{\bf A}^T{\bf S}  + {\bf U}_M^T {\bf W} {\bf N} \label{eq:linear1}
\end{equation}
where ${\bf D} = [ {\bf d}({\bf s}_1), \dots, {\bf d}({\bf s}_N) ] - {\bf u} {\bf 1}_N^T$ and ${\bf N} = [{\bf n}({\bf s}_1),\cdots,{\bf n}({\bf s}_N)]$ are both $M \times N$ matrices. The approximation in 
(\ref{eq:variance_approx}) allows this stacking by using a common whitening matrix ${\bf W}$ for all the sensors. In addition, the covariance matrix of $\mathrm{vec}({\bf U}_M^T {\bf W}{\bf N})$ will be approximately ${\bf I}_{(M-1)N}$.

\subsection{Sensor topology on the Stiefel manifold}

A  Stiefel manifold~\cite{Stiefel1999} in three dimensions, commonly denoted by $\mathcal{V}_{3,3}$, is the set of all $3 \times 3$ unitary matrices ${\bf Q} = [{\bf q}_1,{\bf q}_2,{\bf q}_3] \in \mathbb{R}^{3 \times 3}$, i.e., $\mathcal{V}_{3,3}= \{{\bf Q} \in \mathbb{R}^{3 \times 3} : {\bf Q}^T{\bf Q} = {\bf I}_3\}$. The absolute position of the $n$th sensor can be written as an affine function of a point on the Stiefel manifold, i.e.,
\begin{eqnarray}
{\bf s}_n &=&  c_{n,1}{\bf q}_1 + c_{n,2}{\bf q}_2 + c_{n,3}{\bf q}_3 + {\bf t}\nonumber\\
&=&  {\bf Q} {\bf c}_n + {\bf t}\label{eq:sensorstief}
\end{eqnarray}
where ${\bf t} \in \mathbb{R}^{3 \times 1}$ denotes the translation and is unknown. Note that the combining weights ${\bf c}_n$ are equal to the known coordinates of the $n$th sensor in the reference frame, as introduced in Section \ref{sec:model}. This means that the unknown unitary matrix ${\bf Q}$ actually tells us how the rigid body has rotated in the reference frame.

We can further stack (\ref{eq:sensorstief}) for all the sensors as
\vspace*{-3mm}
\begin{equation}
\vspace*{-3mm}
\small
{\bf S}  =  {\bf Q}{\bf C} + {\bf t}{\bf 1}_N^T= \overbrace{\left[\begin{array}{c|c}{\bf Q} & {\bf t}\end{array}\right]}^{{\bf Q}_e} \overbrace{\left[\begin{array}{c}{\bf C} \\ \hline {\bf 1}^T_N \end{array}\right]}^{{\bf C}_e}.\label{eq:sensorstief1}
\end{equation}
Note that in (\ref{eq:sensorstief1}), we express the unknown sensor locations ${\bf S}$ in terms of the unknown rotations ${\bf Q}$ of a known sensor topology ${\bf C}$ and an unknown translation ${\bf t}$.

\vspace*{-3mm}
\section{The proposed estimators}
In this paper, we consider the novel problem to localize the rigid body by estimating the rotations ${\bf Q}$ and translations ${\bf t}$ in each dimension relative to the anchors. 
The matrix ${\bf Q}$ forms an orthonormal basis for the subspace spanned by the rigid body which reveals all the rotations.
\vspace*{-3mm}\subsection{LS estimator (Unconstrained)} 
Combining (\ref{eq:linear1}) and (\ref{eq:sensorstief1}) results in the following linear model
\begin{equation}
\label{eq:jointLS}
{\bf U}_M^T {\bf W} {\bf D}= -2{\bf U}_M^T{\bf W}{\bf A}^T{{\bf Q}_e}{{\bf C}_e}  + {\bf U}_M^T {\bf W}{\bf N} 
\end{equation}
which can be written as
\begin{equation}
\vspace*{-2mm}
\label{eq:jointLS1}
\bar{\bf D}= \bar{\bf A}{{\bf Q}_e}{{\bf C}_e}  + \bar{\bf N} 
\end{equation}
where $\bar{\bf D} \triangleq {\bf U}_M^T {\bf W}{\bf D} \in \mathbb{R}^{(M-1) \times N}$, $\bar{\bf A} \triangleq -2{\bf U}_M^T{\bf W}{\bf A}^T \in \mathbb{R}^{(M-1) \times 3}$, and $\bar{\bf N}={\bf U}_M^T {\bf W}{\bf N} \in \mathbb{R}^{(M-1) \times N}$. We can further vectorize (\ref{eq:jointLS1}) as
\begin{equation}
\label{eq:jointLS2}
\bar{\bf d}= ({\bf C}_e^T \otimes \bar{\bf A}) {{\bf q}_e}  + \bar{\bf n} 
\end{equation}
where ${{\bf q}_e} = \mathrm{vec}({{\bf Q}_e}) = [{\bf q}_1^T, {\bf q}_2^T, {\bf q}_3^T, {\bf t}^T]^T \in \mathbb{R}^{12 \times 1}$, $\bar{\bf d} = \mathrm{vec}(\bar{\bf D}) \in \mathbb{R}^{(M-1)N \times 1}$, and $\bar{\bf n} = \mathrm{vec}(\bar{\bf N}) \in \mathbb{R}^{(M-1)N \times 1}$. 

To jointly estimate the unknown rotations ${\bf Q}$ and the translations ${\bf t}$ we propose the following joint LS estimator
\begin{equation}
\label{eq:jointLSsolution}
\hat{\mathbf q}_{e,LS}= ({\bf C}_e^T \otimes \bar{\bf A})^\dag \bar{\bf d}
\end{equation} which will have a unique solution if ${\bf C}_e^T \otimes \bar{\bf A}$ has full column-rank which requires $(M-1)N \geq 12$. Finally, we have $\hat{\mathbf Q}_{e,LS} = \mathrm{vec}^{-1}(\hat{\mathbf q}_{e,LS})=\left[\begin{array}{c|c}{\hat{\bf Q}}_{LS} & \hat{\bf t}_{LS}\end{array}\right]$. Note that since $\bar{\bf n}$ is approximately white, we do not use any weighting in the LS formulation.

\subsection{Unitarily constrained LS estimator (CLS)}
The solution of the unconstrained LS estimator does not necessarily lie in the set $\mathcal{V}_{3,3}$, i.e., the columns of the LS estimate $\hat{\bf Q}_{LS}$ obtained in (\ref{eq:jointLSsolution}) are generally not orthogonal to each other and they do not have a unit norm.  

We next propose a LS estimator with a unitary constraint on ${\bf Q}$. For this purpose, we decouple the rotations and the translations in (\ref{eq:sensorstief1}). For this purpose, we adopt a unitary decomposition of ${\bf P}_N \triangleq {\bf I}_N - \frac{1}{N} {\bf 1}_N{\bf 1}_N^T$, i.e.,  ${\bf P}_N = {\bf U}_N{\bf U}_N^T$ where ${\bf U}_N$ is a $N \times (N-1)$ matrix obtained by collecting orthonormal basis vectors of the null-space of ${\bf 1}_N$ so that ${\bf 1}_N^T{\bf U}_N= {\bf 0}$.
Right-multiplying ${\bf U}_N$ to both sides of (\ref{eq:sensorstief1}) leads to
\begin{equation}
{\bf S}{\bf U}_N  =  {\bf Q}{\bf C}{\bf U}_N .\label{eq:sensorstief2}
\end{equation}
Combining (\ref{eq:linear1}) and (\ref{eq:sensorstief2}) we get the following linear model 
\begin{equation}
{\bf U}_M^T {\bf W}{\bf D}{\bf U}_N= -2{\bf U}_M^T{\bf W}{\bf A}^T{\bf Q}{\bf C}{\bf U}_N+ {\bf U}_M^T {\bf W}{\bf N}{\bf U}_N 
\end{equation}
which can be written as
\begin{equation}
\label{eq:model1}
\tilde{\bf D}= \bar{\bf A}{\bf Q}\bar{\bf C}+ \tilde {\bf N}
\end{equation}
where $\tilde{\bf D}= {\bf U}_M^T{\bf W} {\bf D}{\bf U}_N$, $\bar{\bf A} = -2{\bf U}_M^T{\bf W}{\bf A}^T$, $\bar{\bf C} \triangleq {\bf C} {\bf U}_N$, and  $\tilde{\bf N}= {\bf U}_M^T {\bf W}{\bf N}{\bf U}_N$. As before, the covariance matrix of $\mathrm{vec}(\tilde{\bf N})$ will be approximately ${\bf I}_{(M-1)(N-1)}$.

To estimate ${\bf Q}$ we propose a LS problem with a quadratic equality constraint as given by
\begin{equation}
 \label{eq:optproblem}
  \begin{aligned}
&\min_{{\bf Q}} \quad {\|{\bf Q}\bar{\bf C} - {\bf X} \|}_F^2, \\
 &s.t. \quad \quad {\bf Q}^T{\bf Q} = {\bf I}_3
\end{aligned}
\end{equation} where ${\bf X} \triangleq \bar{\bf A}^\dag \tilde{\bf D}$ assuming that $\bar{\bf A}$ has full column-rank.
The optimization problem in (\ref{eq:optproblem}) is non-convex due to the quadratic equality constraint and is commonly referred to as the {\it orthogonal Procrustes problem} (OPP)~\cite{golub1996matrix}.

\begin{myrem} [Anchor positioning]
For $M \geq 3$, the anchor positions can be designed such that the matrix $\bar{\bf A}$ will be full column-rank and well-conditioned. Then, the matrix $\bar{\bf A}$ is left-invertible, i.e., $\bar{\bf A}^\dag \bar{\bf A} = {\bf I}_3$. 
\end{myrem}

 \begin{mylem}[Solution to unitarily constrained LS~\cite{golub1996matrix}] 
The constrained LS problem in (\ref{eq:optproblem}) has an analytical solution $\hat{\bf Q}_{CLS} = {\bf V}{\bf U}^T$ where ${\bf U}$ and ${\bf V}$ are obtained from the singular value decomposition (SVD) of $\bar{\bf C}{\bf X}^T$ which is given by ${\bf U}{\bf \Sigma}{\bf V}^T$. 
\end{mylem}

Subsequently, the LS estimate of the translations ${\bf t}$ can be computed by using $\hat{\bf Q}_{CLS}$ in (\ref{eq:sensorstief1}
) and (\ref{eq:jointLS1}) as 
\begin{equation}
\hat{\bf t}_{CLS} = \frac{1}{N}(\bar{\bf A}^\dag\bar{\bf D}-\hat{\bf Q}_{CLS}{\bf C}){\bf 1}_N.
\end{equation}


\begin{myrem} [Weighted orthogonal Procrustes problem~\cite{WOPP}]
The pseudo inverse operation in (\ref{eq:optproblem}) would color the noise. This can be avoided by solving a weighted orthogonal Procrustes problem. This does not have a closed-form solution, but can be solved using Newton iterations~\cite{WOPP}.
\end{myrem}

\subsection{Unitarily constrained TLS estimator (CTLS)}
In the previous section, we assumed that the sensors are mounted on a rigid body and their topology is accurately known. In practice, there is no reason to believe that errors are restricted only to the range measurements and there are no perturbations on the initial sensor positions. 
The perturbations can be introduced during fabrication of the rigid body or if the body is not entirely rigid.

The position of the $n$th sensor in the reference frame ${\bf c}_n$ is noisy. We denote the perturbation on ${\bf c}_n$ as ${\boldsymbol \delta}_n$, and the perturbations on $\bar{\bf C}$ as ${\boldsymbol \Delta} \triangleq [{\boldsymbol \delta}_1,{\boldsymbol \delta}_2,\ldots,{\boldsymbol \delta}_N]{\bf U}_N$.
Taking the perturbations of the sensor into account we can re-write the data model in (\ref{eq:model1}) as 
\begin{equation}
{\bf Q}(\bar{\bf C} + {\boldsymbol \Delta})  = {\bf X} +  {\bf  E}\label{eq:TLSmodel}
\end{equation}
where $ {\bf  E} \triangleq \bar{\bf  A}^\dag \tilde{\bf  N}$ and ${\bf Q}$ is to be determined as earlier.

The solution to the data model in (\ref{eq:TLSmodel}) leads to the classical TLS optimization problem but with a unitary constraint. The unitarily constrained TLS optimization problem is given by
\begin{equation}
 \label{eq:TLSoptproblem}
  \footnotesize
 \begin{aligned}
&\min_{{\bf Q}} \quad {\|{\boldsymbol \Delta}\|}^2_F + {\|{\bf E}\|}^2_F, \\
&s.t. \quad {\bf Q}(\bar{\bf C} + {\boldsymbol \Delta})  = {\bf X} +  {\bf  E} , \\
& \quad \quad \, \, \, {\bf Q}^T{\bf Q} = {\bf I}_3.
\end{aligned}
\end{equation} 

 \begin{mylem}[Solution to unitarily constrained TLS~\cite{ArunCTLS}] 
The unitarily constrained TLS problem in (\ref{eq:TLSoptproblem}) has the same solution as the unitarily constrained LS problem, and the solution is $\hat{\bf Q}_{CTLS} = {\bf V} {\bf U}^T$.
\end{mylem}

The algorithms to compute the solution for a unitarily constrained LS and TLS are summarized as Algorithm \ref{alg1}.
\begin{algorithm}
\small
\caption{\footnotesize {Unitarily constrained LS or TLS}}
\label{alg1}
\begin{algorithmic}
\REQUIRE $\bar{\bf C}{\bf X}^T$, ${\bf X} = \bar{\bf A}^\dag \tilde{\bf D}$.
\ENSURE SVD $\bar{\bf C}{\bf X}^T = {\bf U}{\bf \Sigma}{\bf V}^T$\\
\quad \quad \quad \quad  $\hat{\bf Q} = {\bf V} {\bf U}^T$\\
\quad \quad \quad \quad  $\hat{\bf t} = \frac{1}{N}(\bar{\bf A}^\dag\bar{\bf D}-\hat{\bf Q}{\bf C}){\bf 1}_N$
\end{algorithmic}
\end{algorithm} \vspace*{-3mm}


\section{Numerical results}
We consider six sensors mounted on a rigid pyramid as shown in Fig.~\ref{fig:model}. The coordinates of the sensors in the reference frame are chosen such that, 
\begin{equation}
\footnotesize
 \label{eq:exampleC}
 \begin{aligned}
{\bf C} = \left[\begin{array}{cccccc}1 & 6 & 7 & 6 & 2 & 2.5 \\0 & 0 & 5 & 5 & 5 & 2.5 \\0 & 0 & 0 & 0 & 0 & 5\end{array}\right]
\end{aligned}
\end{equation} and $M=10$ anchors are deployed uniformly at random in a range of $100 \, \mathrm{m}$. We use a rotation of  $(20, -25, 10) \, \mathrm{degrees}$ in each dimension, which determines ${\bf Q}$, and a translation of $5 \, \mathrm{m}$ along each dimension. The simulations are averaged over $N_{exp} = 1000$ independent Monte-Carlo experiments.

We analyze the performance of the three proposed estimators: 1) LS (unconstrained), 2) unitarily constrained LS, and 3) unitarily constrained TLS. The performance of the estimators for estimating the rotations are provided in terms of the {\it mean angular error} defined as $\frac{1}{3N_{exp}}\sum_{i=1}^{N_{exp}} \sum_{m=1}^{3} \cos^{-1} \left(\frac{{\bf q}_m^{T} \hat{\bf q}_m^{(i)}}{{\|\hat{\bf q}_m^{(n)}\|}_2}\right)$. This is shown in Fig.~\ref{fig:MSE1}. The root mean square error (RMSE) for estimating the corresponding translations $\sqrt{\frac{1}{N_{exp}}\sum_{i=1}^{N_{exp}} {\|\hat{\bf t}^{(i)} - {\bf t}\|}_2^2}$ is shown in Fig.~\ref{fig:MSE2}. Here, $\hat{\bf q}_m^{(i)},m=1,2,3$ and $\hat{\bf t}^{(i)}$ are the parameters estimated during the $i$th Monte-Carlo experiment. Note that in case of the unconstrained LS estimator ${\|\hat{\bf q}_m^{(i)}\|}_2 \neq 1$. 


Simulations are provided for various reference ranges defined as $10 \log_{10} \frac{N_s\kappa SNR}{3c^2} \,\, \mathrm{dB}$. A reference range of $100 \, \mathrm{dB}$ means that for a range of $100 \, \mathrm{m}$ the standard deviation on the estimated range is $1 \, \mathrm{mm}$. The range measurements in (\ref{eq:range}) are corrupted with an i.i.d. Gaussian random process of variance $\sigma^2_m(e_{m,n})$ derived for the corresponding reference range. 
\begin{figure} [!t]
\centering
\subfloat[\footnotesize without perturbations.] {\label{fig:Qp} \includegraphics[height=1.3in,width=3in]{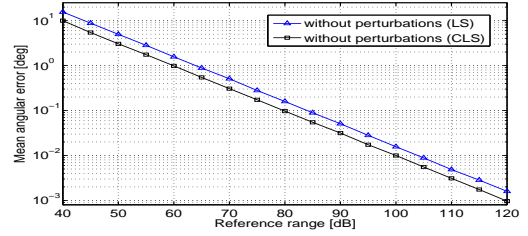}}   \vspace*{-0.7mm}
\subfloat[\footnotesize with perturbations.] {\label{fig:Q} \includegraphics[height=1.3in,width=3in]{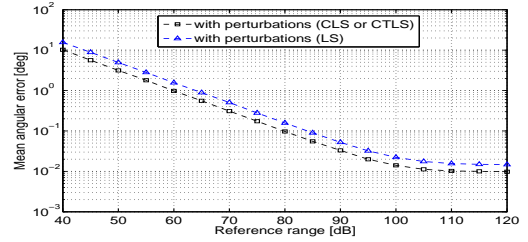}}   \vspace*{-0.7mm}
\caption{\small Mean angular error of the estimated rotations ${\bf Q}$.} \label{fig:MSE1}
\vspace*{-4mm}
\end{figure}
\begin{figure} [!h]
\centering
\subfloat[\footnotesize without perturbations.] {\label{fig:tp} \includegraphics[height=1.3in,width=3in]{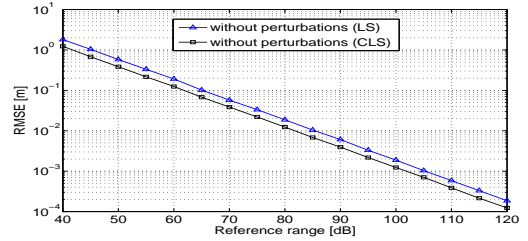}}  \vspace*{-0.7mm}
\subfloat[\footnotesize with perturbations.] {\label{fig:t} \includegraphics[height=1.3in,width=3in]{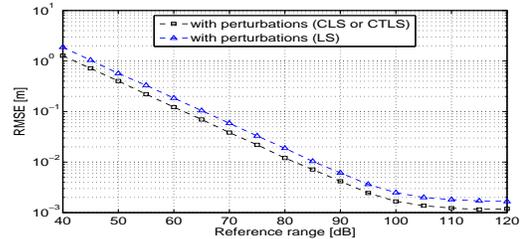}}  \vspace*{-0.7mm}
\caption{\small RMSE of the estimated translations ${\bf t}$.} \label{fig:MSE2}
\vspace*{-5mm}
\end{figure}

For the perturbed case, the sensor topology is corrupted with a zero mean i.i.d. Gaussian process with a standard deviation of  $1 \, \mathrm{mm}$.  The performance of the unconstrained and  constrained estimators in case of perturbations is shown in Fig.~\ref{fig:Q} and Fig.~\ref{fig:t}. 
\vspace*{-3mm}
\section{Conclusions}
We have proposed a problem called rigid body localization, in which the aim is to jointly localize and estimate the orientation of a rigid body in a 3-dimensional space. For rigid body localization, we make use of a few anchors and a known sensor topology of sensors that are mounted on the rigid body. We parameterize the Stiefel manifold using the known sensor topology and propose unconstrained and constrained LS estimators. In order to take the perturbations of the sensor into account, we also propose a unitarily constrained TLS estimator. Incidentally, the solutions to both the constrained LS and constrained TLS estimators are the same. Analytical closed-form solutions for all the estimators have been provided. 

\vfill
\pagebreak

\bibliographystyle{IEEEbib}
\bibliography{IEEEabrv,strings,refs}

\begin{thebibliography}{10}

\bibitem{Gusta05SPM}
F.~Gustafsson and F.~Gunnarsson,
\newblock ``Mobile positioning using wireless networks: possibilities and
  fundamental limitations based on available wireless network measurements,''
\newblock {\em {IEEE} Signal Process. Mag.}, vol. 22, no. 4, pp. 41 -- 53, Jul.
  2005.

\bibitem{localizationSPM}
N.~Patwari, J.N. Ash, S.~Kyperountas, III Hero, A.O., R.L. Moses, and N.S.
  Correal,
\newblock ``Locating the nodes: cooperative localization in wireless sensor
  networks,''
\newblock {\em {IEEE} Signal Process. Mag.}, vol. 22, no. 4, pp. 54 -- 69, Jul.
  2005.

\bibitem{MDS}
Z.-X. Chen, H.-W. Wei, Q.~Wan, S.-F. Ye, and W.-L.ZYang,
\newblock ``A supplement to multidimensional scaling framework for mobile
  location: A unified view,''
\newblock {\em {IEEE} Trans. Signal Process.}, vol. 57, no. 5, pp. 2030 --2034,
  May 2009.

\bibitem{MDShadi}
H.~Jamali-Rad and G.~Leus,
\newblock ``Dynamic multidimensional scaling for low-complexity mobile network
  tracking,''
\newblock {\em {IEEE} Trans. Signal Process.}, vol. 60, no. 8, pp. 4485 --4491,
  Aug. 2012.

\bibitem{underwaterAV}
A.~Caiti, A.~Garulli, F.~Livide, and D.~Prattichizzo,
\newblock ``Localization of autonomous underwater vehicles by floating acoustic
  buoys: a set-membership approach,''
\newblock {\em {IEEE} J. Ocean. Eng.}, vol. 30, no. 1, pp. 140 -- 152, Jan.
  2005.

\bibitem{olfar}
M.J. Bentum, C.J.~M. Verhoeven, A.~J. Boonstra, E.~K.~A. Gill, and A.-J.
  van~der Veen,
\newblock ``A novel astronomical application for formation flying small
  satellites,''
\newblock in {\em Proc. of 60th International Astronautical Congress}, Daejeon,
  October 2009, pp. 1--8, Press IAC.

\bibitem{GPSattitude}
J.-C. Juang and G.-S. Huang,
\newblock ``Development of gps-based attitude determination algorithms,''
\newblock {\em {IEEE} Trans. Aerosp. Electron. Syst.}, vol. 33, no. 3, pp. 968
  --976, Jul. 1997.

\bibitem{accelerometer}
L.~Salhuana,
\newblock ``Tilt sensing using linear accelerometers,''
\newblock in {\em Appl. note AN3461}. February 2012, Freescale Semiconductor.

\bibitem{Stiefel1999}
L.~Eld\'en and H.~Park,
\newblock ``A {P}rocrustes problem on the {S}tiefel manifold,''
\newblock {\em Numerische Mathematik}, vol. 82, pp. 599--619, 1999,
\newblock 10.1007/s002110050432.

\bibitem{ChepuriSPL}
S.~P. Chepuri, R.~Rajan, G.~Leus, and A.-J. van~der Veen,
\newblock ``Joint clock synchronization and ranging: Asymmetrical time-stamping
  and passive listening,''
\newblock {\em {IEEE} Signal Process. Lett.}, vol. 20, no. 1, pp. 51--54, Jan.
  2013.

\bibitem{yiyineurasip}
Y.~Wang and G.~Leus,
\newblock ``Reference-free time-based localization for an asynchronous
  target,''
\newblock {\em EURASIP Journal on Advances in Signal Processing}, vol. 2012,
  no. 1, pp. 19, 2012.

\bibitem{golub1996matrix}
G.H. Golub and C.F. Van~Loan,
\newblock {\em Matrix Computations},
\newblock Johns Hopkins Studies in the Mathematical Sciences. Johns Hopkins
  University Press, 1996.

\bibitem{WOPP}
T.~Viklands,
\newblock {\em Algorithms for the Weighted Orthogonal Procrustes problem and
  Other Least Squares Problems},
\newblock Ph.D. dissertation, Dep. Comput. Sci., Umea Univ., Umea, Sweden.,
  2008.

\bibitem{ArunCTLS}
K.~Arun,
\newblock ``A unitarily constrained total least squares problem in signal
  processing,''
\newblock {\em SIAM Journal on Matrix Analysis and Applications}, vol. 13, no.
  3, pp. 729--745, 1992.

\end{thebibliography}

\end{document}